\documentclass[prl,aps,amsmath,amssymb,amsfonts,twocolumn,nofootinbib,longbibliography,superscriptaddress]{revtex4-2}
\usepackage{amssymb}
\usepackage{bm,mathrsfs}
\usepackage{graphicx}
\usepackage{epsfig}
\usepackage{amsmath,bbm}
\usepackage{amsfonts,amssymb}
\usepackage{times}
\usepackage{verbatim}
\usepackage[sort&compress]{natbib}
\usepackage{amsmath}
\usepackage{bm}
\usepackage{float}

\allowdisplaybreaks[4]
\usepackage[colorlinks,breaklinks,linkcolor=blue,anchorcolor=blue,citecolor=blue,urlcolor=blue]{hyperref}

\usepackage{booktabs}

\begin{document}

\title{Tensor Network Framework for Forecasting Nonlinear and Chaotic Dynamics}

\author{Jia-Bin You}
\email{you\_jiabin@ihpc.a-star.edu.sg}
\affiliation{Quantum Innovation Centre (Q.InC), Agency for Science, Technology and Research (A*STAR), 2 Fusionopolis Way, Innovis \#08-03, Singapore 138634, Republic of Singapore}
\affiliation{Institute of High Performance Computing (IHPC), Agency for Science, Technology and Research (A*STAR), 1 Fusionopolis Way, \#16-16 Connexis, Singapore 138632, Republic of Singapore}

\author{Jian Feng Kong}
\email{kong\_jian\_feng@ihpc.a-star.edu.sg}
\affiliation{Quantum Innovation Centre (Q.InC), Agency for Science, Technology and Research (A*STAR), 2 Fusionopolis Way, Innovis \#08-03, Singapore 138634, Republic of Singapore}
\affiliation{Institute of High Performance Computing (IHPC), Agency for Science, Technology and Research (A*STAR), 1 Fusionopolis Way, \#16-16 Connexis, Singapore 138632, Republic of Singapore}

\author{Jun Ye}
\email{yej@ihpc.a-star.edu.sg}
\affiliation{Quantum Innovation Centre (Q.InC), Agency for Science, Technology and Research (A*STAR), 2 Fusionopolis Way, Innovis \#08-03, Singapore 138634, Republic of Singapore}
\affiliation{Institute of High Performance Computing (IHPC), Agency for Science, Technology and Research (A*STAR), 1 Fusionopolis Way, \#16-16 Connexis, Singapore 138632, Republic of Singapore}

\begin{abstract}
We present a tensor network model (TNM) for forecasting nonlinear and chaotic dynamics, bridging quantum many-body methods with classical complex systems. The TNM leverages hierarchical tensor contractions to encode non-Markovian temporal correlations and multiscale structures, enabling compact and interpretable representations of chaotic flows. Using the Lorenz and R\"{o}ssler systems as benchmarks, we show that the TNM accurately reconstructs short-term trajectories and faithfully captures the attractor geometry.
The model enables robust short-term forecasting beyond several Lyapunov times, offering a meaningful horizon for data-driven prediction under chaos.
Inhomogeneous parametrization of weight tensors improves convergence and robustness compared to homogeneous parametrization, while scaling with bond dimension reveals saturation beyond modest values, consistent with the low intrinsic dimensionality of the chaotic attractor. This work establishes tensor networks as a universal paradigm for data-driven modeling of complex dynamical systems, offering physically motivated control of model expressivity and opening pathways toward applications in climate systems and hybrid quantum-classical simulations.
\end{abstract}

\maketitle

\paragraph{Introduction.}
Accurately forecasting chaotic dynamics remains a central challenge across physics, climate science, and engineering. Chaos induces exponential divergence of trajectories, limiting long-horizon predictability and complicating data-driven modeling. In particular, chaotic attractors exhibit non-Markovian temporal correlations and multiscale structures that conventional approaches often struggle to capture.

Tensor networks (TNs), originally developed for quantum many-body physics, provide compact low-rank representations with tunable expressivity through the bond dimension. Prominent architectures include matrix product states (MPS) \cite{Schollwoeck2011}, projected entangled pair states (PEPS) \cite{VerstraeteCirac2004}, multi-scale entanglement renormalization ansatz (MERA) \cite{Vidal2008}, and tree tensor networks (TTN) \cite{TagliacozzoEvenblyVidal2009}. These structures decompose complex correlations into interconnected local tensors, significantly reducing computational complexity while preserving essential features. The hierarchical contraction structure of TNs mirrors renormalization flows in statistical physics, successively integrating short-range fluctuations into effective long-range behavior. Beyond quantum systems, TNs have inspired applications in machine learning, pattern recognition, and the analysis of nonlinear dynamics \cite{StoudenmireSchwab2016,ChenEtAl2018,GlasserEtAl2018}.

The Lorenz system provides a canonical benchmark for nonlinear dynamics and chaos \cite{Lorenz1963,Strogatz2018}. Originally introduced to model atmospheric convection, it exhibits strange attractors, exponential sensitivity to initial conditions, and complex multiscale structure. The R\"{o}ssler system \cite{ROSSLER1976397}, introduced as another paradigmatic chaotic flow, displays qualitatively distinct attractor geometry and dynamical features. Together, the Lorenz and R\"{o}ssler systems provide representative testbeds that capture diverse aspects of chaotic behavior, making them well-suited for exploring new forecasting approaches.

Machine learning methods, including recurrent neural networks, long short-term memory (LSTM) models \cite{Shahi2022,Vlachas2018,Fan2020}, and physics-informed neural networks (PINNs) \cite{WANG2024115326,DOAN2020101237}, have been explored for chaotic forecasting. While effective in some regimes, these approaches often require large parameter counts and lack interpretability, motivating the search for compact and physically grounded alternatives.

In this work, we propose a tensor network model (TNM) for forecasting chaotic dynamics, demonstrated on the Lorenz and R\"{o}ssler attractors. The TNM leverages hierarchical tensor contractions to efficiently encode non-Markovian temporal correlations, providing a natural way to represent the multiscale memory structure of chaotic flows, where short-term sensitivity, intermediate recurrences, and long-term attractor constraints coexist. This hierarchical structure naturally mirrors the multiscale nature of chaotic dynamics, making TNM particularly well-suited for such systems. This compact representation enables the model to reproduce short-term trajectories with high accuracy and to faithfully reconstruct the underlying dynamics.
Notably, the TNM supports consistent forecasting across several Lyapunov times, offering a practically relevant predictive horizon despite the intrinsic instability of chaotic systems.

Beyond demonstrating predictive capability, we analyze the role of bond dimension \( D \) as a physically motivated measure of expressivity. Increasing \( D \) systematically improves prediction accuracy until saturation, revealing that only a limited number of effective degrees of freedom govern the underlying dynamics. These chaotic attractors, while embedded in three-dimensional phase space, in fact resides on a manifold of intrinsically low dimensionality, and this structure is faithfully captured by modest bond dimension. Furthermore, we investigate parameterization strategies and find that inhomogeneous weight tensors accelerate convergence and yield more robust performance than their homogeneous counterparts.

These results establish tensor networks as a versatile and interpretable paradigm for chaotic forecasting. By bridging methods from quantum many-body physics with nonlinear dynamics, the TNM provides compact surrogates with clear scaling properties. This approach opens pathways toward applications in climate systems, where memory effects are critical, and in hybrid quantum-classical simulations, where compact and scalable representations are critical for practical applications.

\begin{figure}
\centering
\includegraphics[width=\columnwidth]{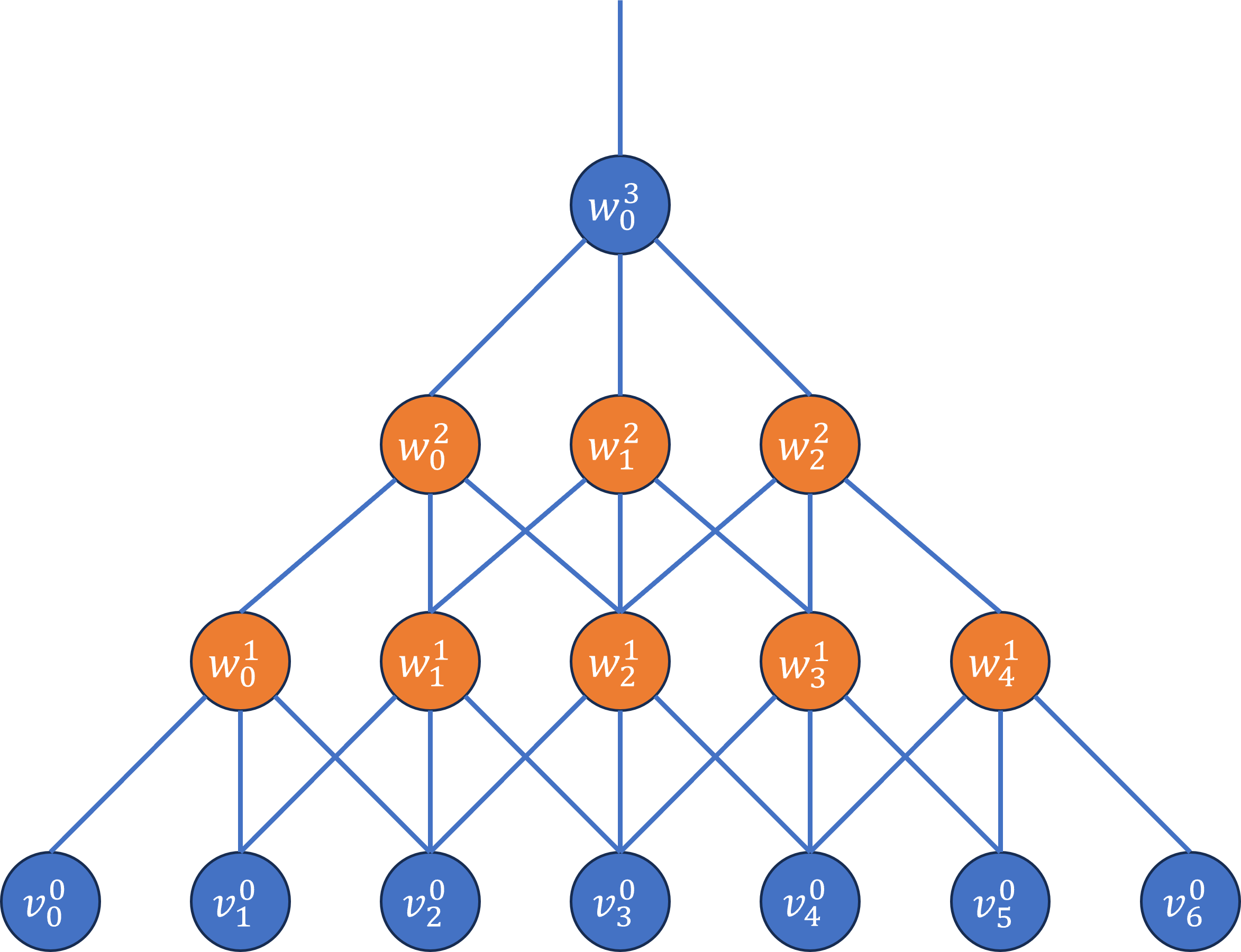}
\caption{Schematic of the tensor network model (TNM), which maps a sequence of past states to the next state via hierarchical tensor contractions, efficiently encoding non-Markovian correlations in chaotic dynamics. The bottom layer (\( v_0^0, v_1^0, \dots, v_6^0 \)) represents input vectors at successive time steps, each with feature dimension \( d \). Hidden layers consist of rank-4 weight tensors performing tensor contractions to extract hierarchical correlations. The output node at the top encodes the predicted future state.}
\label{fig1}
\end{figure}

\paragraph{Tensor network model.}
Fig.~\ref{fig1} illustrates the hierarchical architecture of the TNM employed in this work. Each circle denotes a tensor node, with the bottom layer (\( v^0_0, v^0_1, \dots, v^0_6 \)) corresponding to input vectors at successive time steps, each of dimension \( d \). The intermediate hidden layers (orange circles, e.g., \(w^1_i\) and \(w^2_i\)) perform tensor contractions that aggregate local features into higher-level abstract representations. In the forward pass, each hidden node carries out a multilinear update,
\begin{equation}
\begin{split}
v_j^{l}[m] = f\left(\sum_{n,o,p} w_{j}^{l}[m, n, o, p] v^{l-1}_{j}[n] v^{l-1}_{j+1}[o] v^{l-1}_{j+2}[p] \right),\\
\end{split}
\label{contraction}
\end{equation}
where \(f(\cdot)\) is the nonlinear activation, and \(w_{j}^{l}[m, n, o, p]\) is a learnable rank-4 weight tensor. Depending on the layer, the weight tensor may take the form \(D \times d \times d \times d\), \(D \times D \times D \times D\), or \(d \times D \times D \times D\). The bond dimension \( D \) serves as a physically motivated parameter that controls the expressive power of the TNM, analogous to the role of entanglement truncation in quantum many-body simulations.

At each level \(l\), a hidden node combines three adjacent nodes from level (\(l-1\)), performing the contraction in Eq.~(\ref{contraction}) to generate a higher-level feature vector. This recursive procedure constructs a hierarchy of representations that naturally encodes multi-way correlations and long-range temporal dependencies, yielding a compact yet expressive description of the dynamics. Such a structure provides an interpretable way to model non-Markovian memory effects, while maintaining efficiency in parameter count.

The hierarchical contraction culminates at the top node, which outputs the predicted future state. Explicitly, the final layer computes
\begin{equation}
\begin{split}
v^3_0[m] = \sum_{n,o,p} w_{0}^3[m, n, o, p] v_{0}^2[n] v_{1}^2[o] v_{2}^2[p],\\
\end{split}
\label{output}
\end{equation}
with no activation applied at this stage, thereby yielding the model output. This formulation highlights the ability of TNMs to combine compact representation, hierarchical correlation extraction, and physically interpretable scaling with bond dimension, making them naturally suited for modeling nonlinear chaotic dynamics.

\paragraph{Benchmark systems.}
To assess the performance of the TNM, we adopt the Lorenz system as a canonical benchmark for chaotic dynamics. Introduced to describe simplified thermal convection, the Lorenz model has become a paradigmatic example in nonlinear dynamics and chaos theory. It is governed by three coupled nonlinear ordinary differential equations: \( \dot{x} = \sigma(y - x) \), \( \dot{y} = x(\rho - z) - y \), and \( \dot{z} = xy - \beta z \), where \( x, y, z \) denote fluid velocity, temperature gradients, and convective intensity, respectively. The dynamics are controlled by three positive parameters: the Prandtl number \(\sigma\), the Rayleigh number \(\rho\), and a geometric factor \(\beta\). As \(\rho\) increases, the system undergoes transitions from fixed-point stability (\(\rho < 1\)) to periodic oscillations (\(1 < \rho < 24.74\)) and ultimately to fully developed chaos (\(\rho > 24.74\)). The standard chaotic regime arises at \(\sigma = 10\), \(\beta = 8/3\), and \(\rho = 28\), yielding the well-known Lorenz attractor. Additional results for the R\"{o}ssler system are provided in the Supplemental Material \cite{SM} to illustrate the generality of the approach.

\begin{figure*}
\includegraphics[width=1\linewidth]{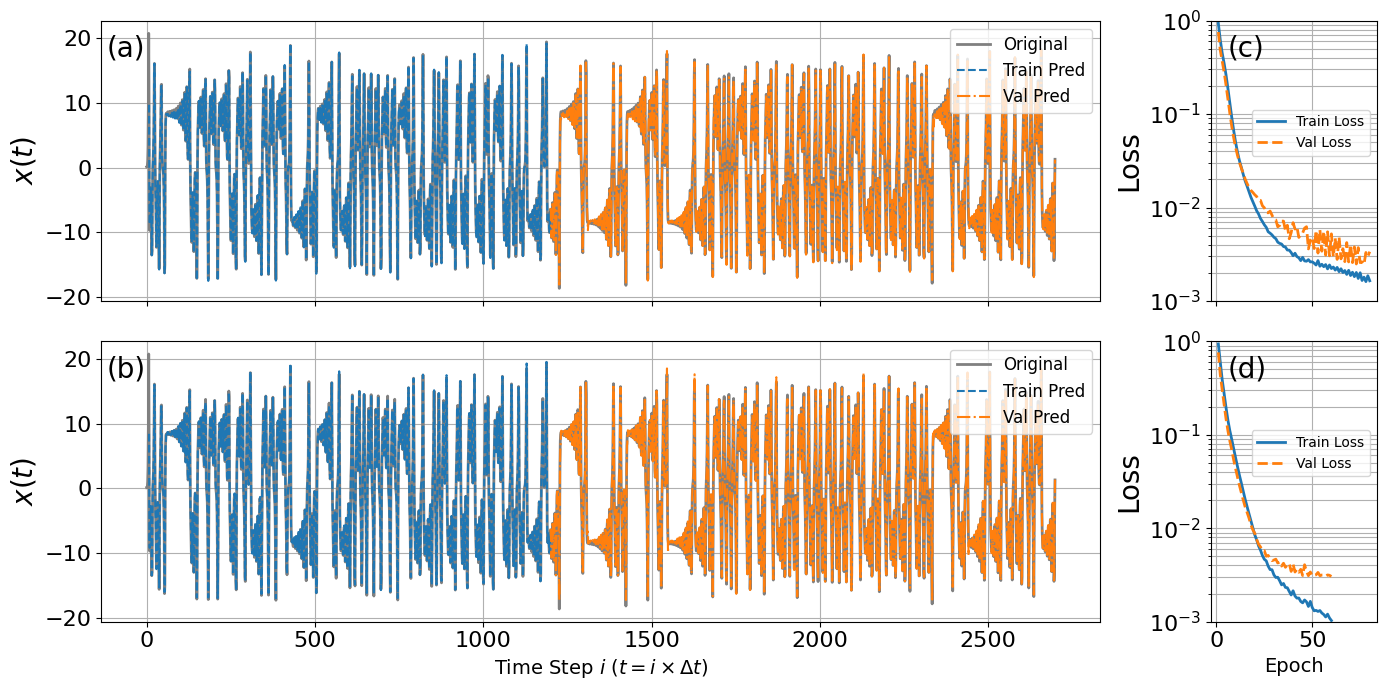}
\caption{Forecasting performance of the TNM on the Lorenz system with homogeneous (a),(c) and inhomogeneous (b),(d) parametrizations of weight tensors. Panels (a),(b) show the evolution of \( x(t) \): gray solid curves denote the true trajectory, blue dashed lines are training predictions, and orange dash-dot lines are validation predictions. Both models reproduce the dynamics, but the inhomogeneous case adheres more closely to the ground truth. Panels (c),(d) display training and validation losses on a logarithmic scale, showing faster convergence and lower final errors for the inhomogeneous model. The bond dimension is fixed at 
\( D=8 \), and training is performed by the Adam optimizer with a learning rate of 0.001, using 80 epochs for the homogeneous and 60 epochs for the inhomogeneous model.}
\label{fig2}
\end{figure*}

\begin{figure}
\includegraphics[width=1\linewidth]{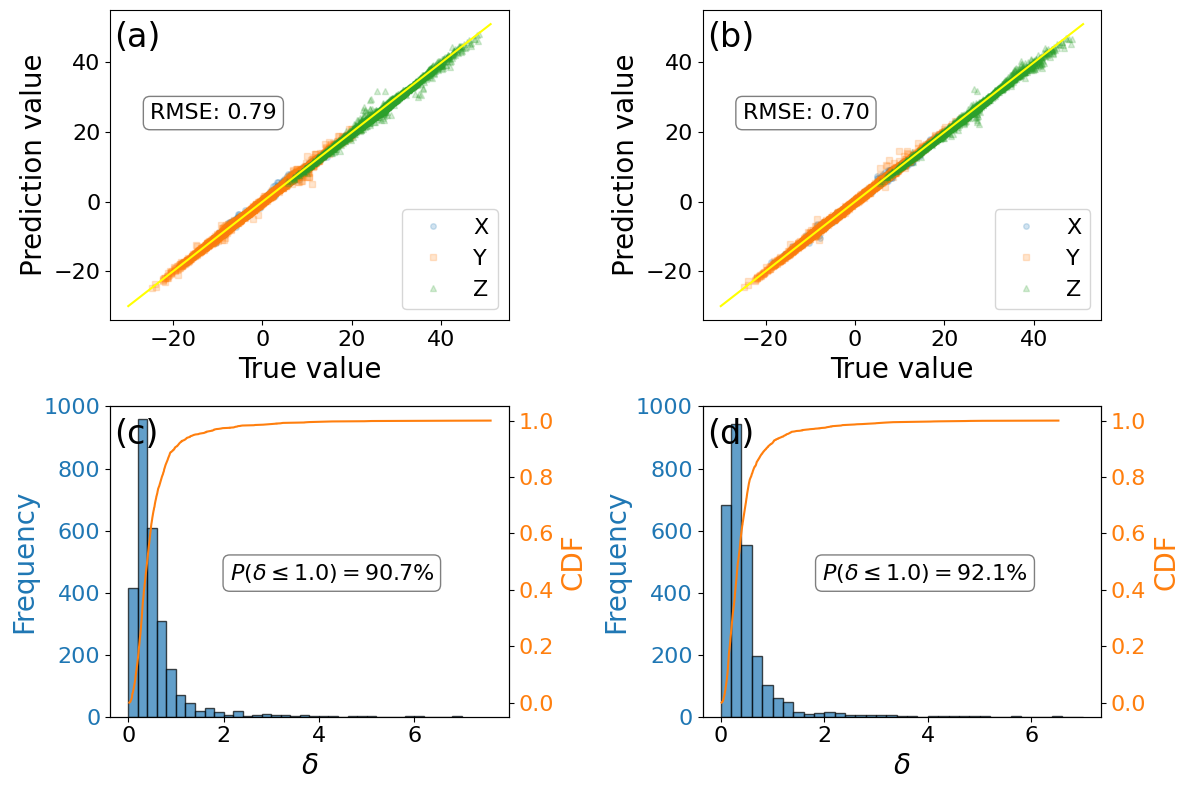}
\caption{Prediction performance of TNM on the Lorenz attractor.
(a) Parity plot for the homogeneous model, showing predicted versus true coordinates \((x, y, z)\) with RMSE = 0.79.
(b) Parity plot for the inhomogeneous model, achieving improved accuracy with RMSE = 0.70.
(c) Histogram and cumulative distribution function (CDF) of pointwise Euclidean prediction error for the homogeneous model. 90.7\% of predictions lie within a distance of 1.0 from the ground truth.
(d) Same as (c) but for the inhomogeneous model, where 92.1\% of predictions fall below the error threshold of 1.0.
Together, these results demonstrate the superior predictive accuracy and robustness of the inhomogeneous TNM.}
\label{fig3}
\end{figure}

\paragraph{Numerical implementation.}
For training and evaluation, we numerically integrate the Lorenz equations with a step size of 0.01. To construct the dataset, system states are recorded every 10 integration steps, corresponding to an effective sampling interval of \( \Delta t = 0.1 \). This yields a sequence of phase-space vectors \( \{v_{0}^{0}, v_{1}^{0}, \dots, v_{N-1}^{0}\} \), where each \( v_{i}^{0} = \{ x_i, y_i, z_i \} \) specifies the system state at time \(t = i \times 0.1\). The TNM processes these trajectories by mapping a window of past states to the next one, formally approximating the non-Markovian relation \( v_{7}^{0} = f_{\text{TNM}}(v_{0}^{0}, v_{1}^{0}, \dots, v_{6}^{0}) \), as illustrated in Fig.~\ref{fig1}.

A total of \( N = 3000 \) samples are generated and divided into training (40\%), validation (50\%), and test (10\%) subsets. Each feature dimension of the training and validation data is standardized using the mean and variance from the training set, ensuring consistency and preventing data leakage during optimization.

The TNM parameters are trained by backpropagation to minimize the mean squared error (MSE) between predicted and true trajectories. We consider two parameterization schemes: in the homogeneous case, all weight tensors within a hidden layer are constrained to be identical, \( w_{j}^{i} = w_{k}^{i} \), whereas in the inhomogeneous case, each weight tensor is treated independently, offering greater flexibility. Model parameters are updated using the Adam optimizer, with validation error monitored after each epoch to track generalization.

After convergence, the trained TNM is employed to generate predictions both on historical data (training and validation sets) and on unseen test data. Future evolution is obtained recursively, with model outputs fed back as new inputs, thereby demonstrating the forecasting capability of the network.

\paragraph{Trajectory reconstruction and accuracy.}
Fig.~\ref{fig2} demonstrate the performance of the TNM in forecasting Lorenz dynamics and compare homogeneous with inhomogeneous parametrizations of weight tensors. As shown in Fig.~\ref{fig2}(a) and (b), both variants reproduce training and validation trajectories with good fidelity, closely tracking the true evolution of $x(t)$. Fig.~\ref{fig2}(c) and (d) show the training and validation losses on a logarithmic scale. The inhomogeneous model converges more rapidly and reaches lower final errors than the homogeneous case, indicating that increased parameter flexibility not only improves expressivity but also stabilizes optimization. Together, these results highlight that inhomogeneous parametrization enhances both learning efficiency and predictive generalization in chaotic time-series forecasting.

Predictive accuracy is quantified in Fig.~\ref{fig3}. The inhomogeneous model achieves a lower root-mean-square error, $\mathrm{RMSE} = \sqrt{N^{-1}\sum_{i=1}^{N}\|\vec r_i^{\,\text{pred}}-\vec r_i^{\,\text{true}}\|^2}$ with $\vec r_i=(x_i,y_i,z_i)$, yielding 0.70 compared to 0.79 for the homogeneous case, and its parity plot shows tighter alignment with the identity line (yellow solid line). The error statistics are further resolved into the empirical distribution of Euclidean errors, $\delta_i = \|\vec r_i^{\,\text{pred}}-\vec r_i^{\,\text{true}}\|$, and their cumulative distribution functions (CDFs). The histograms reveal that the inhomogeneous model concentrates predictions in the low-error region, producing a narrower peak and suppressed long tails, while the homogeneous case exhibits broader spread and more frequent large deviations. The CDFs quantify this contrast: $P(\delta \leq 1.0)$ reaches 92.1\% for the inhomogeneous case versus 90.7\% for the homogeneous one.

These findings highlight not only the improvement in average accuracy but also enhanced robustness against chaotic trajectories. This advantage originates from the hierarchical contraction structure of the TNM, which mirrors the multiscale nature of chaotic dynamics by successively coarse-graining local correlations. The faster convergence of the inhomogeneous parametrization can be understood as an adaptive allocation of representational capacity, where each tensor specializes to different features of the trajectory, thereby improving efficiency while retaining interpretability.

\begin{figure}
\includegraphics[width=1\linewidth]{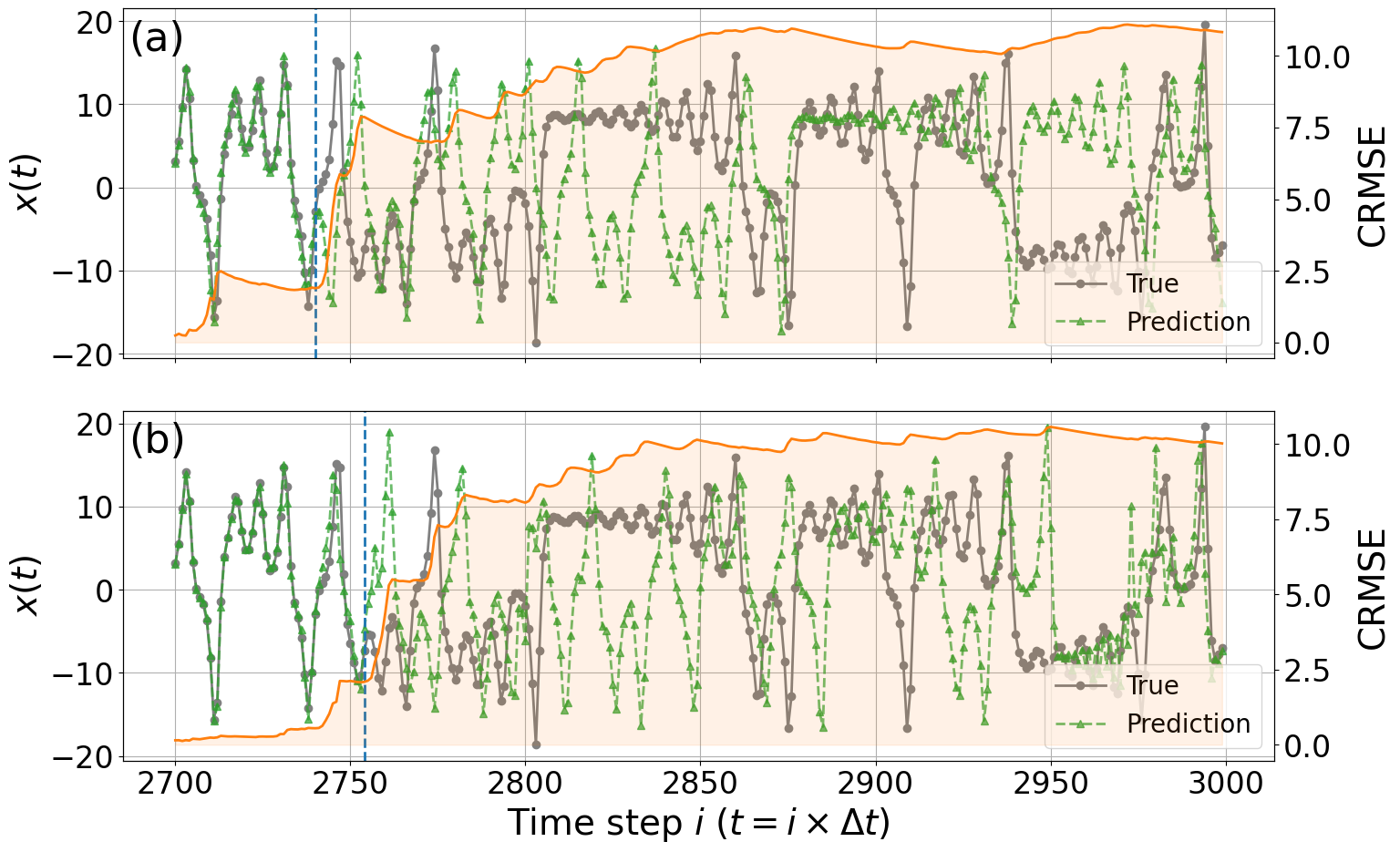}
\caption{Forecasting performance of the TNM on unseen test data using homogeneous (a) and inhomogeneous (b) parametrizations. Gray soild curves show the ground-truth trajectory \( x(t) \), green dashed lines indicate TNM predictions, and orange lines (right axis) represent cumulative RMSE (CRMSE). Bond dimension is set to \( D=8 \), and models are trained with Adam optimizer at learning rate 0.001.}
\label{fig4}
\end{figure}

\begin{figure}
\includegraphics[width=0.8\linewidth]{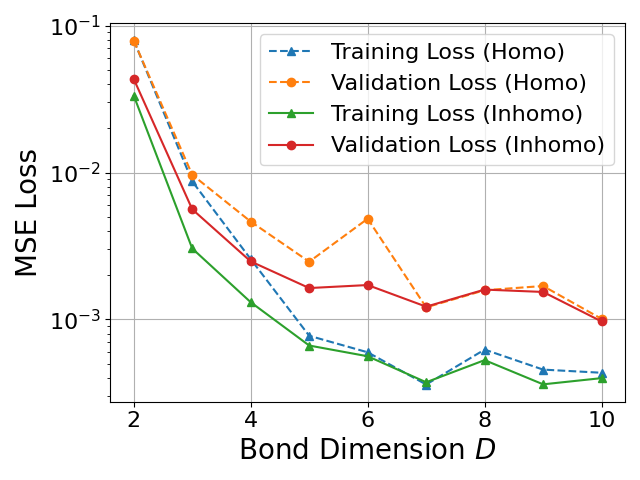}
\caption{Dependence of TNM performance on the bond dimension \( D \), comparing homogeneous (dashed lines) and inhomogeneous (solid lines) parametrizations. Training (triangles) and validation (circles) losses are shown at epoch 200. Increasing \( D \) from 2 to 5 reduces the losses for both cases, but further gains plateau for \( D>5 \). At all bond dimensions, the inhomogeneous parametrization achieves consistently lower losses than the homogeneous one.}
\label{fig5}
\end{figure}

\paragraph{Forecast horizon and Lyapunov stability.}
Fig.~\ref{fig4} evaluates the generalization capability of the TNM on unseen test data using a recursive prediction strategy where predicted outputs are iteratively fed back as inputs. Both the homogeneous and inhomogeneous parametrizations faithfully reproduce the qualitative features of the Lorenz trajectory over short timescales, demonstrating the ability of TNM to extrapolate beyond the training and validation regimes.

To quantify the predictive horizon, we examine the cumulative RMSE (CRMSE) growth over time. The Lorenz system, under the chosen parameters, has a largest Lyapunov exponent \( \lambda_1 \approx 0.9056 \) \cite{Sprott2003}, corresponding to a Lyapunov time of \( T_\lambda \approx 1.1 \). With time step \( \Delta t = 0.1 \), one Lyapunov time spans approximately 11 steps. As shown in Fig.~\ref{fig4}, the homogeneous model maintains CRMSE \( < 1.9 \) for about 40 steps (\( \sim 3.6 T_\lambda \), blue dashed line), while the inhomogeneous model extends this to around 54 steps (\( \sim 5.0 T_\lambda \), CRMSE \( < 2.1 \), blue dashed line). This improvement highlights the increased representational flexibility of inhomogeneous parametrization in capturing multi-scale temporal correlations. While long-horizon predictability remains fundamentally constrained by the exponential divergence inherent to chaotic trajectories, the TNM achieves stable forecasting across several Lyapunov times, constituting a meaningful benchmark for data-driven modeling of nonlinear systems.

\paragraph{Bond-dimension scaling.}
Fig.~\ref{fig5} shows the influence of the bond dimension \( D \) by comparing homogeneous and inhomogeneous parametrizations of the TNM. As \( D \) increases from 2 to 5, both losses decrease markedly, reflecting improved modeling accuracy as the model gains capacity to represent richer temporal correlations. For \( D>5 \), however, the performance gains saturate and the losses reach a plateau, suggesting that additional model complexity yields only marginal improvements. Across all bond dimensions, the inhomogeneous parametrization achieves consistently lower losses than the homogeneous counterpart, confirming its greater expressivity and learning efficiency.

These results emphasize the role of bond dimension as a physically motivated parameter that balances capacity with generalization. They further demonstrate that the Lorenz dynamics can be faithfully captured within a compact, low-dimensional representation. The observed saturation for \( D>5 \) indicates that the dynamics can be reliably captured with modest bond dimension, highlighting that additional complexity yields diminishing returns. This is in line with the low intrinsic complexity of the Lorenz attractor, whose fractal dimension is approximately 2.06 \cite{PhysRevLett.50.346,GRASSBERGER1983189}. These results demonstrate that seemingly complex chaotic dynamics can, in fact, be efficiently captured by compact TNM with modest bond dimension.

\paragraph{Conclusions.}
We have introduced a tensor network model to forecast the Lorenz and R\"{o}ssler systems, which are canonical examples of chaotic dynamics. By exploiting the hierarchical and multilinear structure of tensor networks, the TNM efficiently encodes non-Markovian temporal correlations and captures multiscale features of chaotic flows. Numerical experiments demonstrate accurate reconstruction of trajectories and predictive performance on training, validation and unseen test data, with inhomogeneous parametrization yielding faster convergence and improved robustness compared to homogeneous parametrization. Exploration of bond dimension reveals systematic improvements up to \( D\approx5 \), followed by saturation. The TNM maintains consistent predictive accuracy across several Lyapunov times, providing a meaningful horizon for short-term forecasting despite the exponential sensitivity inherent to chaotic dynamics.

These findings highlight tensor networks as a compact and interpretable paradigm for chaotic forecasting, providing a physically grounded control parameter, the bond dimension, to balance model capacity with generalization. While demonstrated here on the Lorenz and R\"{o}ssler systems, the TNM framework suggests potential applicability to more complex dynamical systems, such as partial differential equations and hybrid quantum-classical simulations. Our results point to tensor networks as a promising tool for advancing data-driven modeling of complex systems.

\paragraph{Acknowledgments.}
This research is supported by A*STAR under A*STAR (C230917003 and C230917007), and Q.InC Strategic Research and Translational Thrust.

\clearpage
\onecolumngrid

\begin{center}
  \textbf{\large Supplemental Material for\\[4pt]
  Tensor Network Framework for Forecasting Nonlinear and Chaotic Dynamics}\\[6pt]
\end{center}

\setcounter{section}{0}
\renewcommand{\thesection}{S\arabic{section}}
\renewcommand{\thefigure}{S\arabic{figure}}
\renewcommand{\thetable}{S\arabic{table}}
\setcounter{figure}{0}
\setcounter{table}{0}

This Supplemental Material provides additional results for the R\"ossler system, complementing the main text where the Lorenz system is used as the primary benchmark.
We first summarize the governing equations and parameter choices of the R\"ossler system.
We then report three sets of simulations: (i) trajectory reconstruction and training dynamics under homogeneous vs.\ inhomogeneous parametrizations; (ii) quantitative accuracy via parity plots and empirical error distributions; and (iii) generalization under recursive forecasting, together with cumulative error growth.

\section{R\"ossler system}
As a complementary benchmark to the Lorenz system, we consider the R\"ossler system
\begin{equation}
\dot{x} = -y - z,\qquad
\dot{y} = x + a\,y,\qquad
\dot{z} = b + z(x-c),
\label{eq:Rossler}
\end{equation}
With parameters \(a=0.2\), \(b=0.2\), \(c=5.7\), which yield a canonical chaotic regime characterized by a spiral-type attractor \cite{Sprott2003}.
Trajectories are generated by numerically integrating Eq.~(\ref{eq:Rossler}) with a fixed step size \(\Delta t=0.01\), and uniformly sampled every 10 integration steps, corresponding to an effective sampling interval \(\Delta t_{\text{sample}}=0.1\), to construct input--target pairs for training and validation of the tensor network model (TNM).
For accuracy metrics we use the root-mean-square error
\(
\mathrm{RMSE}=\sqrt{N^{-1}\sum_{i=1}^{N}\|\vec r_i^{\,\text{pred}}-\vec r_i^{\,\text{true}}\|^2}
\),
and the pointwise Euclidean error
\(
\delta_i=\|\vec r_i^{\,\text{pred}}-\vec r_i^{\,\text{true}}\|
\),
with \(\vec r_i=(x_i,y_i,z_i)\).

\section{Trajectory reconstruction}
\begin{figure}
  \centering
  \includegraphics[width=0.8\linewidth]{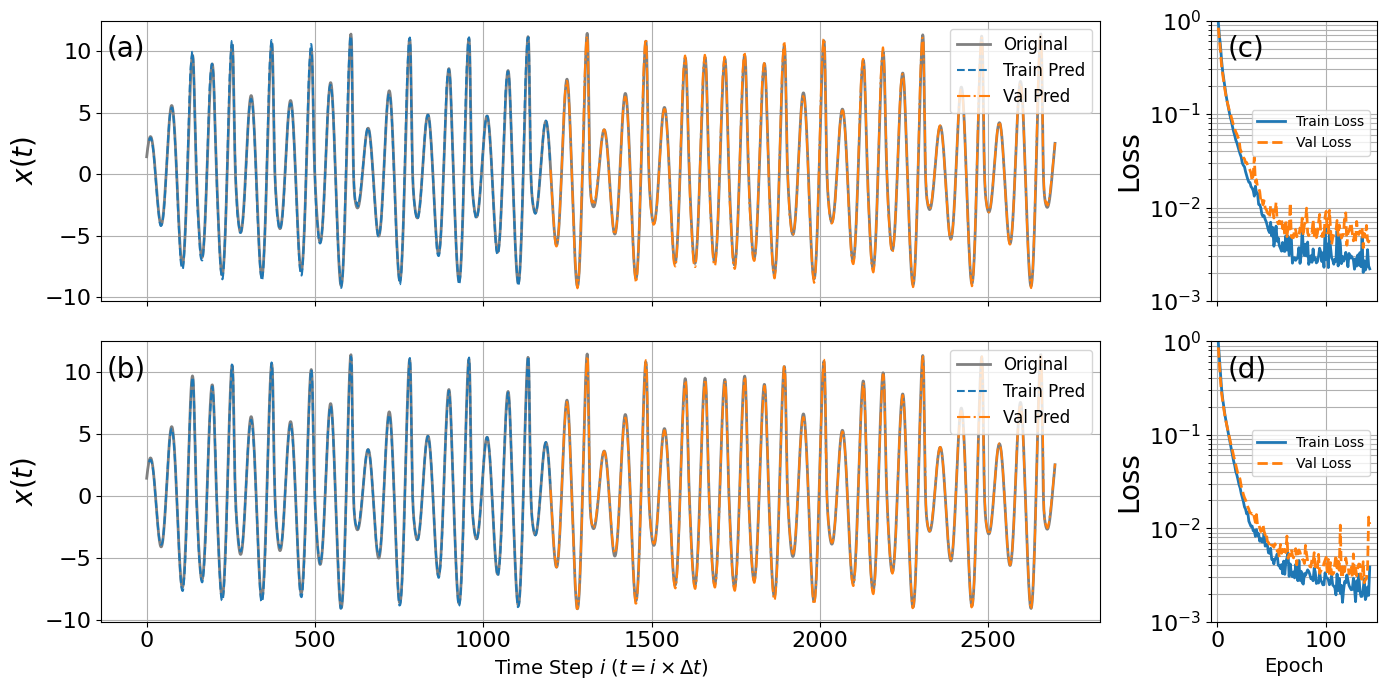}
  \caption{Trajectory reconstruction and training dynamics for R\"ossler attractor.
  (a),(b) Predictions of \(x(t)\) on training (blue dashed) and validation (orange dash-dot) segments compared with the true trajectory (gray). 
  (c),(d) Training and validation losses on a logarithmic scale. 
  Both parametrizations reproduce the dynamics with high fidelity and converge rapidly to loss values of order \( 10^{-3}\sim10^{-2} \).}
  \label{fig:S1}
\end{figure}

Fig.~\ref{fig:S1} compares homogeneous and inhomogeneous parametrizations of weight tensors in the TNM for the R\"ossler system. 
Panels (a),(b) show predictions of \(x(t)\) on the training and validation segments, which both closely reproduce the true trajectory. 
Panels (c),(d) display the training and validation losses on a logarithmic scale.
The training and validation losses drop rapidly within the first few dozen epochs to the \( 10^{-3}\sim10^{-2} \) range. By epoch 140, the validation losses of the homogeneous and inhomogeneous cases are both around \( 3\times10^{-3} \).

\section{Prediction accuracy}
\begin{figure}
  \centering
  \includegraphics[width=0.5\linewidth]{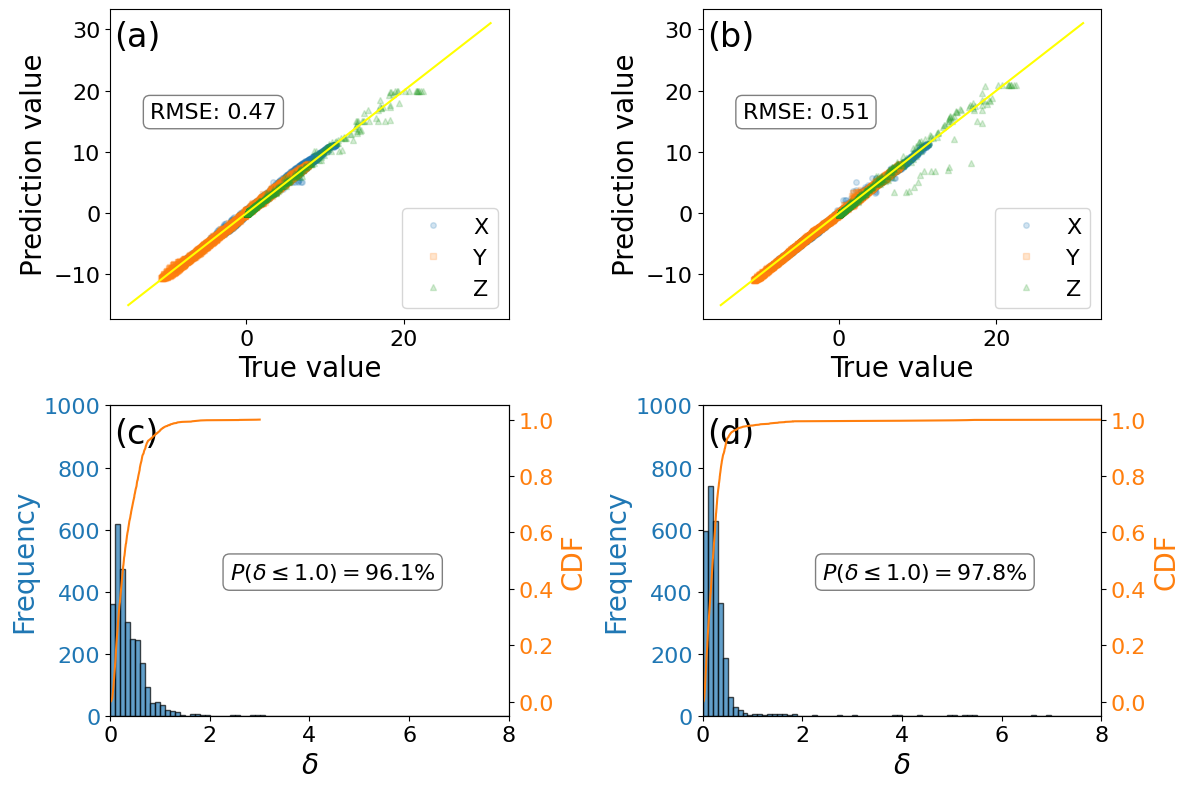}
  \caption{Quantitative accuracy of the TNM on the R\"{o}ssler system.
  (a),(b) Parity plots of prediction versus true values for \( x,y,z \), with RMSE annotated. The homogeneous parametrization attains a lower overall RMSE (0.47 vs.\ 0.51).
(c),(d) Empirical error distributions and cumulative distribution functions (CDF, orange). The inhomogeneous parametrization yields a higher proportion of small-error predictions (\( P(\delta\le1.0)=97.8\% \)) compared to the homogeneous case (96.1\%).
The results indicate comparable performance, with the homogeneous parametrization reducing average error and the inhomogeneous case achieving a tighter bulk distribution.}
  \label{fig:S2}
\end{figure}

Fig.~\ref{fig:S2} evaluates the prediction accuracy of the TNM on the R\"{o}ssler system. Parity plots for the three coordinates [panels (a),(b)] show that both parametrizations reproduce the true values with high fidelity. The homogeneous parametrization yields a lower root-mean-square error (RMSE = 0.47) compared to the inhomogeneous case (RMSE = 0.51). However, error distributions [panels (c),(d)] reveal a complementary trend: the inhomogeneous parametrization achieves a higher fraction of small-error predictions, with \( P(\delta\le1.0)=97.8\% \), compared to 96.1\% for the homogeneous model. Thus, while the homogeneous parametrization improves the average RMSE, the inhomogeneous parametrization concentrates a larger proportion of predictions near the true trajectory, reflecting a narrower bulk error distribution but occasional larger deviations.

\section{Generalization and recursive forecasting}
\begin{figure}
  \centering
  \includegraphics[width=0.5\linewidth]{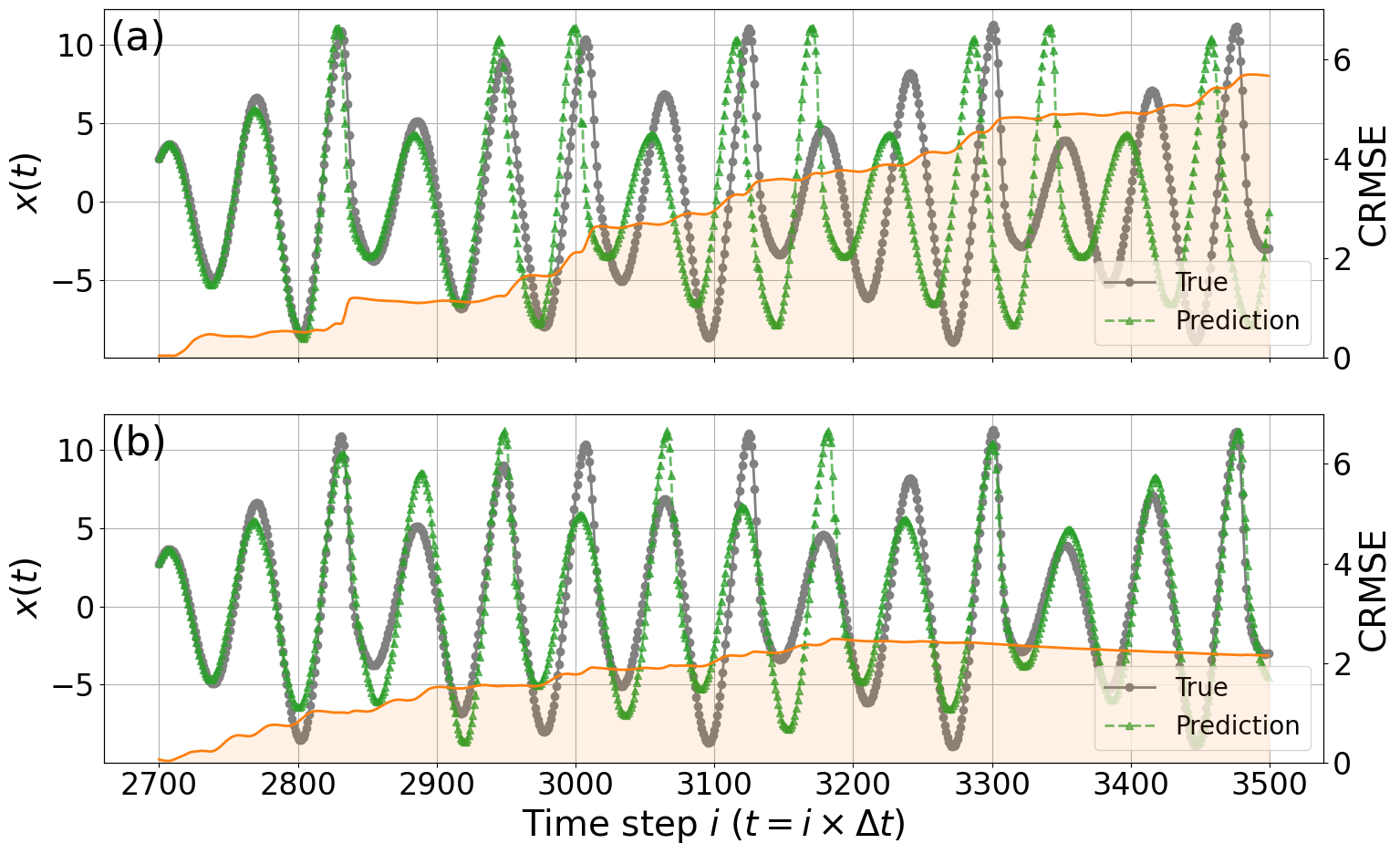}
  \caption{R\"{o}ssler recursive forecasting beyond the training/validation range.
(a),(b) Autoregressive predictions of \( x(t) \) (green) compared with the true trajectory (gray). The orange curve shows the cumulative RMSE (CRMSE; right axis) evaluated over this interval. In (a) the error grows steadily to \(\sim5.5\), reflecting phase and amplitude drift, whereas in (b) the growth is slower and remains near \(\sim2.0\) by the end of the window. Both models capture the short-term oscillatory structure, while the inhomogeneous parametrization exhibits slower error accumulation at longer horizons.}
  \label{fig:S3}
\end{figure}

Figure~\ref{fig:S3} examines generalization under recursive forecasting on the R\"{o}ssler system. After the onset of autonomous prediction, the model output is iteratively fed back as input, and the orange curve reports the cumulative RMSE (CRMSE, right axis) measured from this point. Both parametrizations reproduce the short-term oscillatory structure of \( x(t) \). In panel (a) the phase and amplitude drift accumulate, and CRMSE increases nearly monotonically to \(\sim5.5\) by the end of the window. In panel (b) the trajectory remains phase-aligned for a longer interval. CRMSE grows more slowly and stays bounded around \(\sim2.0\) near the end, with a mild decrease in the final portion. Thus, while long-horizon divergence is expected for chaotic dynamics, the inhomogeneous parametrization exhibits markedly slower error accumulation during autonomous prediction.


\end{document}